\documentclass[12pt,preprint]{aastex}

\renewcommand{\cite}{\citealp}

\newcommand{\rrl}{{RR~Lyrae}}

\shorttitle{Variable stars in the Bo${\rm \ddot{o}}$ dSph}
\shortauthors{Dall'Ora et al.}

\begin{document}

\title{Variable stars in the newly discovered Milky Way satellite in Bo${\rm \ddot{o}}$tes
\altaffilmark{1}}

\author{
Massimo Dall'Ora,\altaffilmark{2}
Gisella Clementini,\altaffilmark{3}
Karen Kinemuchi, \altaffilmark{4}
Vincenzo Ripepi,\altaffilmark{2}
Marcella Marconi, \altaffilmark{2}
Luca Di Fabrizio,\altaffilmark{5}
Claudia Greco, \altaffilmark{3} 
Christopher T. Rodgers, \altaffilmark{4} 
Charles Kuehn, \altaffilmark{6} 
Horace A. Smith, \altaffilmark{6}
}

\altaffiltext{1}{Based on data collected at the 1.52 m telescope of the
INAF-Osservatorio Astronomico di Bologna, Loiano, Italy, at the 
INAF-Telescopio Nazionale Galileo, La Palma, Canary Islands, and at the 2.3 m
telescope at the Wyoming Infrared Observatory (WIRO) at Mt. Jelm, Wyoming, USA.}

\altaffiltext{2}{INAF, Osservatorio Astronomico di Capodimonte, 
via Moiarello 16, I-80131 Napoli, Italy,
(dallora, ripepi, marcella)@na.astro.it}

\altaffiltext{3}{INAF, Osservatorio Astronomico di
Bologna, via Ranzani 1, I-40127 
Bologna, Italy;
(gisella.clementini, claudia.greco)@oabo.inaf.it}

\altaffiltext{4}{University of Wyoming, Department of Physics \& Astronomy, Dept. 3905, 
Laramie, WY 82071, USA;
(kinemuch, crodgers)@uwyo.edu}

\altaffiltext{5}{INAF, Centro Galileo Galilei \& Telescopio Nazionale Galileo, PO Box 565, 38700 S.
Cruz de La Palma, Spain; difabrizio@tng.iac.es}

\altaffiltext{6}{Department of Physics and Astronomy, Michigan State University, East Lansing, 
MI 48824-2320, USA;
(kuehn, smith)@pa.msu.edu}

\begin{abstract}

We present $V,I$ light curves for 12 variable stars identified in
the newly discovered satellite of the Milky Way in the Bo${\rm
  \ddot{o}}$tes constellation \citep{belokurov06}.  Our sample
includes 11 RR Lyrae stars (5 first overtone, 5 fundamental mode and
1 double mode pulsator), and one long period variable close to the
galaxy red giant branch tip.  The RR Lyrae stars trace very well the
average $V$ luminosity of the galaxy horizontal branch, leading to a
true distance modulus for the galaxy of $\mu_0$=19.11 $\pm$ 0.08 mag
for an assumed metal abundance of [Fe/H]=$-2.5$ \citep{munoz06}, and
for $E(B-V)$=0.02 mag. Average periods are {\rm $\langle
  Pab\rangle$}=0.69 d and {\rm $\langle Pc\rangle$}=0.37 d for {\it
  ab-} and {\it c-} type RR Lyrae stars, respectively, making of
Bo${\rm \ddot{o}}$tes the second pure Oosterhoff type II (OoII) dSph
after Ursa Minor.  The location of the double mode RR Lyrae (RRd) in
the Petersen diagram is consistent with RRd stars in OoII clusters,
and corresponds to an intrinsic luminosity of $log L/log
L\odot$=1.72 (for Z=10$^{-4}$ and M=0.80 M$\odot$) according to
\citet{bono96} pulsation models.
\end{abstract}

\keywords{
galaxies: dwarf
---galaxies: individual (Bo${\rm \ddot{o}}$tes)
---galaxies: distances and redshifts
---stars: horizontal branch 
---stars: variables: other 
---techniques: photometry
}

\section{Introduction}

The discovery of a new satellite companion of the Milky Way 
was recently announced by \citet{belokurov06}, who named it the  
Bo${\rm \ddot{o}}$tes dwarf spheroidal (dSph) after the constellation of
Bo${\rm \ddot{o}}$tes in which the structure was detected.
\citeauthor{munoz06} (\citeyear{munoz06}) obtained spectroscopy of red and
asymptotic giant branch stars in the Bo${\rm \ddot{o}}$tes dSph galaxy 
from which they measured a very low metal abundance
of [Fe/H]$\sim -2.5$ dex, confirming the results obtained by \citet{belokurov06} based on the 
galaxy color-magnitude diagram (CMD). This low metallicity makes of 
 Bo${\rm \ddot{o}}$tes the most metal-poor Local Group dSph known so far.
The galaxy  $i, g-i$ CMDs
published by \citet{belokurov06} and 
\citet{munoz06} clearly show a horizontal branch (HB) that extends 
to the blue enough to cross the classical instability strip, thus suggesting 
that RR Lyrae stars should be present in the 
Bo${\rm \ddot{o}}$ 
dSph. 
RR Lyrae stars have been found in all the Local Group dSphs where observations were
deep enough to reach the HB. They trace the old stellar population ($t> 10$ Gyr) of the host
system and, being standard candles, provide an invaluable tool to estimate the 
galaxy distance. 

In this Letter we present $V,I$ light curves for 12 variable stars we have identified in the 
Bo${\rm \ddot{o}}$tes 
dSph, periods and pulsation properties for 11 of them, and the first CMD of the galaxy in the $V,I$
bands of the Johnson-Cousins photometric system.  Our sample includes
11 of the variable stars for which \citeauthor{siegel06}
(\citeyear{siegel06}, hereinafter S06) presents $B,I$
photometry and periods evaluated by fitting the observed $B$ light
curves with RR Lyrae templates by \citet{layden98}.

Observations and data reductions are described in Section 2. The color
magnitude diagram of the 
Bo${\rm \ddot{o}}$ 
dSph is presented in Section 3.
Identification and pulsation
characteristics (periods, amplitudes, and types) of the variable stars 
are described in Section 4.
In Section 5 we discuss the distance we derived for this new
satellite of the Milky Way using independent methods based on the RR Lyrae stars. 
Final results are summarized in Section 6.

\section{Observations and data reductions}

Time series $V,I$ photometry of the Bo${\rm \ddot{o}}$tes 
dSph galaxy
has been collected between April and July 2006 with 
BFOSC at the 1.52 m telescope of the Bologna Observatory in 
Loiano\footnote{http://www.bo.astro.it/loiano/index.htm}, and with 
DOLORES at the 3.5 m TNG telescope\footnote{http://www.tng.iac.es/instruments/lrs/}, 
whereas $B,V,I$ observations have been gathered at the 2.3 m Wyoming Infrared 
Observatory (WIRO) telescope, using the WIRO-Prime CCD prime focus camera
\footnote{http://physics.uwyo.edu/$\sim$amonson/wiro/prime.html}. 
In this letter we present results from the analysis of the 
$V,I$ data. A more extended analysis of the full dataset
will be presented in a forthcoming paper. 
Due to the different field of view of the three instrumental set-ups, 
the number of phase points ranges from a minimum of 17 and 4 to a maximum of 52 and 17 
in the $V$ and $I$ bands, respectively. 
Typical internal errors of our $V$ band photometry for single phase points
at the level of the HB are in the range from 0.01 to 0.03 mag.

Images were pre-reduced following standard techniques (bias subtraction and 
flat-field correction) with IRAF\footnote{IRAF is distributed by the National Optical Astronomical 
Observatories, which are operated by the Association of Universities for
Research in Astronomy, Inc., under cooperative agreement with the 
National Science Foundation}, and $I$-band images were 
corrected for fringing by adopting, for each instrument, a well-suited fringing 
map.
We measured the star magnitudes by aperture photometry, since the field is not
crowded and the image quality of a number of frames was not sufficient for a
satisfactory PSF evaluation. After a preliminary PSF reduction performed with the
DaophotII/Allstar packages \citep{stetson87,stetson92}, aperture photometry was
carried out with a radius of 1.5 FWHM for each individual image.

The absolute photometric calibration was obtained during the 
night of May 8, 2006, through the observation at the TNG of standard stars in   
Landolt field PG1633 \citep{landolt92} as extended by P.B. Stetson\footnote{ 
http://cadcwww.dao.nrc.ca/standards/}. 
For the atmospheric extinction coefficient, we adopted the average 
value for La Palma\footnote{see http://www.ast.cam.ac.uk/dwe/SRF/camc\_extinction.html}.
A total number of 67 standard stars covering approximately the 
color interval 0.6$< V-I <$ 2.7 were used to derive the 
calibration equations which have $\sigma=\pm$0.016 and $\pm$0.019 mag scatter, in $V$ and
$I$, respectively.  
The derived color terms are in good agreement with those found on the
DOLORES@TNG web page and in the literature. Since the aperture
photometry of the standard stars was performed with an aperture
radius of 5.5 arcsec, we derived a correction for the (smaller)
aperture used in the two reference scientific frames. This correction
is 0.06 and 0.07 $\pm0.03$ mag in $V$ and $I$, respectively. 
The total uncertainty 
of our photometry is $\sigma_V$=0.035 mag and $\sigma_I$=0.04 mag, in $V$ and $I$, 
respectively. 

\section[]{The galaxy CMD}
The $I,V-I$ CMD of the Bo${\rm \ddot{o}}$tes 
 dSph is shown in Figure 1.  The two lines are the mean ridge lines of
the globular clusters (GCs) M15 (solid line) and M3 (dashed line),
drawn from the cluster's CMDs available at Padua University WEB
site\footnote{http://dipastro.astro.unipd.it/globulars, note that the Padua photometry
is calibrated to the Landolt standard system.}, and shifted
in magnitude and color to fit the galaxy HB. 
The Bo${\rm \ddot{o}}$tes  
red giant branch (RGB) is very well fitted by 
a metal-poor cluster like M15 ([Fe/H]$\sim -2.3$), thus
supporting the 
low metal abundance derived for the
galaxy by \citet{munoz06}. 
A low metal abundance in the range from $\sim -2.0$, to $-2.5$ dex is also 
suggested by the location in the Petersen diagram 
of a double mode pulsator we have
identified in the galaxy, (see Section 5).
If we adopt for the galaxy the distance modulus we infer from the 
average luminosity of the RR
Lyrae stars (see Section 5) and adopt the 
\citet{bella04} calibration of the tip $I$ luminosity,  
we estimate that the galaxy 
RGB tip is at $I$=15.40 mag
for [Fe/H]=$-2.5$.
Bo${\rm \ddot{o}}$tes 
brightest star in our CMD is V16 at 
$V$=15.97 and $I$=14.60, hence about 0.8 mag above the galaxy
tip.  

\section[]{Identification of the variable stars} 

Variable stars were identified from the $V$ band time series 
where we have a larger number of phase points and then 
counter-identified in $I$. 
First we 
calculated the Fourier transform (in \citealt{sc96}
formulation) for
each star in the photometric catalog with at least 12
epochs, then
we averaged this transform to estimate the noise and
calculated the signal-to-noise ratios.
We then checked all the stars with high S/N, that
for the RR Lyrae stars typically went from 25 to 160. 
Coordinates for the variable stars were obtained cross correlating
our catalog with the catalog MAST/casg available at http://www-gsss.stsci.edu.
The r.m.s of the cross correlation is 0.07 arcsec for both coordinates.

All the RR Lyrae stars in our sample are also in the S06 database, while
S06's V1, V7, V11 and V13 are outside our field of view. We were able
as well to assess the variability of the bright star located close to
the tip of the Bo${\rm \ddot{o}}$tes RGB (see Fig. 1) for which S06 does not provide a light
curve. 

Periods were derived using GRaTiS (Graphycal Analyzer of Time Series) a custom  
software developed at the Bologna Observatory (see \citealt{df99,clementini00}).
 We were able to derive reliable periods for 11 of the 12 stars
we identified in the  
Bo${\rm \ddot{o}}$tes dSph. 
Given the about  100 days window 
of our observations these periods are accurate to
the fourth digit.
Along with the star's average luminosity, the light variation amplitude and    
CMD position allowed us to classify the variable star type.
Our sample includes 5 RRc's, 5 RRab's, 1 RRd and 1 long period variable (LPV). 
The properties of the variable stars are summarized in Table 2.
An electronic table of dates and
    magnitudes is available 
by request.

Examples of the light curves are shown in Figure 2, 
where data for the RR Lyrae stars are folded using the periods and epochs 
of maximum light provided in Table ~2.
V16  turned out to be a LPV.
In the 100 days spanned by 
our observations we only observed
a portion of the star's light curve, thus only a rough estimate of the
period could be obtained.
In 
Figure 2 we show for V16 the sequence of observations in HJD.

Our periods agree with S06's periods rounded to the second
digit for 9 out of the 11 variables we have in common. However, the S06 periods,
although similar to ours, do not provide the best fit to our data.
Furthermore, there are two remarkable differences with respect to S06
classification, namely stars: V5 and V12. These stars have been classified by S06 respectively
RRc and RRab. The S06 period for V5 (0.3863158 d) does not phase our data. Our best fit period
for the star is 0.6506 d. A shorter period of about 0.3943 d could possibly phase our data, 
although with a larger scatter and it is thus considered less likely. We also find that the star 
has very small amplitude, on average about 0.2 smaller than 
other {\it c-}type RR Lyrae stars, and is about 0.2 mag brighter than
the galaxy HB. 
Both the reduced amplitude and the overluminosity might be caused by V5 being blended
with a companion star.
%
However, additional and 
better resolution data
would be needed to assess the star actual nature.
Given its peculiarities and the less certain classification in type, in the following
discussion this star will always be considered separately.
The most striking difference with S06 is star V12  that we find to be a double 
mode RR Lyrae star, with periods of $P_1$=0.3948 and $P_0$=0.5296 d, and 
period ratio of 0.7455 d. The double mode behavior of V12 is clearly seen
in Figure 2, and the star's position near the transition between {\it c-} and {\it ab-}
type RR Lyrae stars 
seem to support our finding.

The average periods of the RR Lyrae stars are: 
{\rm $\langle Pc\rangle$}=0.37 d ($\sigma$=0.04, average on 
5 stars; average and $\sigma$ do not change whether we include or not the double mode star) 
and {\rm $\langle Pab\rangle$}=0.69 d ($\sigma$=0.12, average on 
4 stars) or 0.68 ($\sigma$=0.10, average on 
5 stars) if the 
variable star brighter than the HB is considered. These average values and the high
frequency of {\it c-}type stars 
are consistent
with the  
Bo${\rm \ddot{o}}$tes 
dSph being an Oosterhoff type II system (\citealt{oo39}), and along with 
the low metal abundance make of the galaxy an analog of the Ursa Minor dSph galaxy.
The $V$ period-amplitude relation of the 
Bo${\rm \ddot{o}}$tes 
RR Lyrae stars is shown 
in Figure 3 and compared with the distributions of the M3 RRab stars.
The 4 Bo${\rm \ddot{o}}$tes
{\it ab-}type RR Lyrae stars with certain classification (open circles in 
Figure 3) are found to lie close to the region occupied by the  
well-evolved M3 RRab stars (dashed curve in Figure 3, taken from 
\citealt{cacciari05}) and the Oosterhoff type II
locus.
According to Table 1 these stars appear also to be slightly brighter
than the average level of the Bo${\rm \ddot{o}}$tes HB and in Figure 1
lie slightly above the HB ridge lines of M15 and M3, thus 
confirming they could be evolved objects. In conclusion, all the evidence
agrees that the Bo${\rm \ddot{o}}$tes dSph is an old, metal-poor object
of Oosterhoff II class.  
%

\section[]{The distance of the Boo dSph galaxy}

The average luminosity of the Bo${\rm \ddot{o}}$tes
 RR Lyrae stars is 
${\rm \langle  V\rangle}$=19.546 $\pm 0.018$ ($\sigma$=0.057, average on 10 stars) and 
${\rm \langle V\rangle}$=19.531 $\pm 0.022$ ($\sigma$=0.073, average on 11 stars) if the 
overluminous star is included.
Assuming for the absolute luminosity of the RR Lyrae stars at [Fe/H]=$-1.5$
$M_V$=0.59$\pm$0.03  (\citealt{cc03}), $\Delta M_V/[Fe/H]$=0.214 ($\pm$ 0.047)
 mag/dex (\citealt{clementini03}, \citealt{gratton04}) for the slope of the luminosity 
 metallicity relation, 
 $E(B-V)$=0.02 (according to \citealt{belokurov06} and in agreement with the small color shift 
 required to match M15 and
 the Bo${\rm \ddot{o}}$tes 
 horizontal and red giant branches), and [Fe/H]=$-2.5$ (Munoz et al. 2006), 
 the galaxy distance modulus is: 19.11 $\pm$ 0.08 or 19.09 $\pm$ 0.08 if the overluminous 
 variable is included, where errors  include the standard
 deviation of the average, and the uncertainties in the photometry, reddening, and
 RR Lyrae absolute magnitude.   
 
The presence of a double-mode RR Lyrae star in  Boo 
allows us to estimate its mass, hence derive the galaxy distance,
by comparing observed and predicted period ratios in the Petersen diagram
\citep{petersen72} for double mode RR Lyrae stars. \citet{clementini04}
published an updated version of this diagram 
and compared it with the pulsation models of
\citet{bono96} and \citet{bragaglia01}.
The position of V12 in the Petersen diagram is shown in Figure 4 which is an
update of \citet{clementini04} figure 14. The star location in the diagram 
seems consistent with RRd stars in
OoII clusters, and we find the star's mass is close to  0.80 $M\odot$, in
agreement with the evolutionary predictions for HB stars populating
the instability strip at $Z$=0.0001 (see e.g. \citealt{pietri04}).
This mass, combined with the luminosity  provided by the
Petersen diagram and the observed fundamental period of V12, allows us
to evaluate the star's $\log T_e$ from the pulsation equation.
We used Equation 1 in \citet{dc04}. 
Once the mass, luminosity, and effective temperature are known,  
the star's absolute visual magnitude was calculated using the model atmospheres
by \citeauthor{castelli97} (\citeyear{castelli97}a,b). We find 
 $M_V$(V12)=0.52, with a conservative error of $\sim$ 0.13 mag.
The corresponding distance modulus for the  
Bo${\rm \ddot{o}}$tes dSph
is then
$\mu_0=V_0-M_V=19.01 \pm 0.15$ mag.

We estimated the distance to 
the Bo${\rm \ddot{o}}$tes 
system with two further 
independent methods:i) the comparison of theoretical and
empirical RR Lyrae first overtone
blue edges (FOBE) and fundamental red edges (FRE) in the $M_V$
versus $\log P$ plane (see \citealt{caputo97}, \citealt{caputo99}, \citealt{dc04}) and,
 ii) the theoretical Wesenheit function (see \citealt{dc04}). 
>From the first technique we 
find that the \citet{dc04} theoretical boundaries of
the instability strip (FOBE and FRE) for metal poor pulsators 
with mean mass of 0.80 $M_{\odot}$ match the 
observed RR Lyrae stars in 
Bo${\rm \ddot{o}}$tes
for a distance modulus of 19.21$\pm$0.10 masg.
A distance modulus of $19.17 \pm 0.12$ mag is obtained instead 
by adopting the theoretical
Wesenheit relation for the same value of the mass and discarding V5. 
All these independent distance determinations based on the RR Lyrae stars are well consistent to 
each other within their quoted uncertainties.

\section[]{Summary and conclusions}

We have identified and obtained $V,I$ light curves for 11 RR Lyrae stars 
(5 RRc, 5 RRab and 1 RRd) and
one long period variable in the newly discovered dSph galaxy in 
Bo${\rm \ddot{o}}$tes \citep{belokurov06}.
We have investigated the Oosterhoff classification of this
dwarf galaxy with the RR Lyrae stars.  We have found that 
the RR Lyrae parameters support
an Oosterhoff II class for the Bo${\rm \ddot{o}}$tes dSph and that the galaxy
{\it ab-}type RR Lyrae stars might be slightly evolved objects.

>From the average luminosity of the RR Lyrae stars the galaxy distance modulus
is  $\mu_0$=19.11 $\pm$ 0.08 (D=66 $\pm$ 3 kpc). Three further RR Lyrae-based independent
methods confirm this distance within the observational errors, leading to a weighted mean
modulus of  $\mu_0$=19.14 $\pm$ 0.07 mag.

\acknowledgments 
We warmly thank Luciana Federici for providing the mean ridge lines of M15 and M3, 
S. Galleti, G. Andreuzzi and the staff of the Loiano and TNG telescopes. 
Financial support for this study was provided by MIUR, under the scientific
project 2004020323, (P.I.: Massimo Capaccioli). 
HAS thanks the US National Science Foudation for support (AST 0607249).

\clearpage
 


\clearpage

\begin{table}
\scriptsize
\caption[]{Identification and properties of the variable stars in the Bo${\rm \ddot{o}}$tes dSph
      galaxy}
         \label{t:bootes_var}
     $$
         \begin{array}{lcclllcrcrrrc}
	    \hline
            \hline
           \noalign{\smallskip}
           {\rm Name} &  {\rm \alpha } & {\rm \delta} &  {\rm Type} &~~~~{\rm P} & 
	    ~~~{\rm Epoch}  & {\rm \langle V\rangle} & N_V& {\rm \langle I\rangle} & N_I & {\rm A_V}~~ & {\rm A_I}~~ & {\rm Notes}\\
            ~~{\rm (a)}& {\rm (2000)}& {\rm (2000)}& & ~{\rm (days)}& ($-$2450000) &(b)& & (b) & &  &  & \\
            \noalign{\smallskip}
            \hline
            \noalign{\smallskip}
	    
{\rm ~V2} & 13:59:51.34 & 14:39:06.0 & {\rm RRc }    & 0.3119:&3852.729:   & 19.63 & 14 &19.32& 3 &0.34:&>0.13~&\\
{\rm ~V3} & 14:00:26.86 & 14:35:33.1 & {\rm RRc }    & 0.3232 &3852.808:   & 19.58 & 19 &19.25& 3 &0.57~&>0.15~&\\
{\rm ~V4} & 14:00:08.90 & 14:34:24.1 & {\rm RRc }    & 0.3860 &3847.4073   & 19.57 & 42 &19.19& 7 &0.59~& 0.20~&\\
{\rm ~V5} & 14:00:21.56 & 14:37:28.8 & {\rm RRab:}   & 0.6506:&3886.8486   & 19.38 & 15 &18.93& 4 &0.33~&  $-$~~~&{\rm (c)}\\
{\rm ~V6} & 13:59:45.95 & 14:31:40.7 & {\rm RRc }    & 0.3919 &3852.778	   & 19.58 & 50 &19.13& 15&0.53~& 0.22~&\\
{\rm ~V8} & 13:59:59.69 & 14:27:34.0 & {\rm RRc }    & 0.4179 &3852.891	   & 19.54 & 49 &19.10& 17&0.40~& 0.26~&\\
{\rm ~V9} & 13:59:47.28 & 14:27:56.3 & {\rm RRab}    & 0.5755 &3881.4475   & 19.55 & 52 &19.07& 14&1.00~&>0.41~&\\
{\rm ~V10}& 14:00:25.75 & 14:33:08.4 & {\rm RRab}    & 0.6280 &3860.418	   & 19.47 & 27 &18.98& 4 &1.09~&>0.24~ &\\
{\rm ~V12}& 13:59:56.00 & 14:34:55.0 & {\rm RRd }    & 0.3948 &3852.891	   & 19.59 & 41 &19.13& 8 &0.53~&>0.20~& {\rm (d)}\\
{\rm ~V14}& 13:59:25.75 & 14:23:45.3 & {\rm RRab}    & 0.7186 &3846.994:   & 19.50 & 13 &19.03& 2 &0.44:&  $-$~~~ &\\
{\rm ~V15}& 14:00:11.09 & 14:24:19.7 & {\rm RRab}    & 0.8456 &3851.856	   & 19.45 & 41 &18.91& 14&0.48~& 0.28:& \\
{\rm ~V16}& 13:59:48.74 & 14:34:48.2 & {\rm LPV}     & $$\sim 85$$&~~~~~~$-$ & 15.97 & 33 &14.60& 1 &>0.43~&  $-$~~~ &\\
\hline
            \end{array}
	    $$
{\small $^{\mathrm{a}}$ Variable stars from  V2 to V15 coincide with stars in S06.}\\
{\small $^{\mathrm{b}}$ ${\rm \langle V\rangle}$ and ${\rm \langle I\rangle}$ values are 
intensity-averaged mean magnitudes.}\\
{\small $^{\mathrm{c}}$ Variable star about 0.2 mag brighter than the galaxy HB.\\
{\small $^{\mathrm{d}}$ Double mode \rrl\ star with fundamental mode period $P_0$=0.5296 d, and period ratio 
$P_1$/$P_0$=0.7455.}} 
\end{table}

\clearpage

\begin{figure}
\plotone{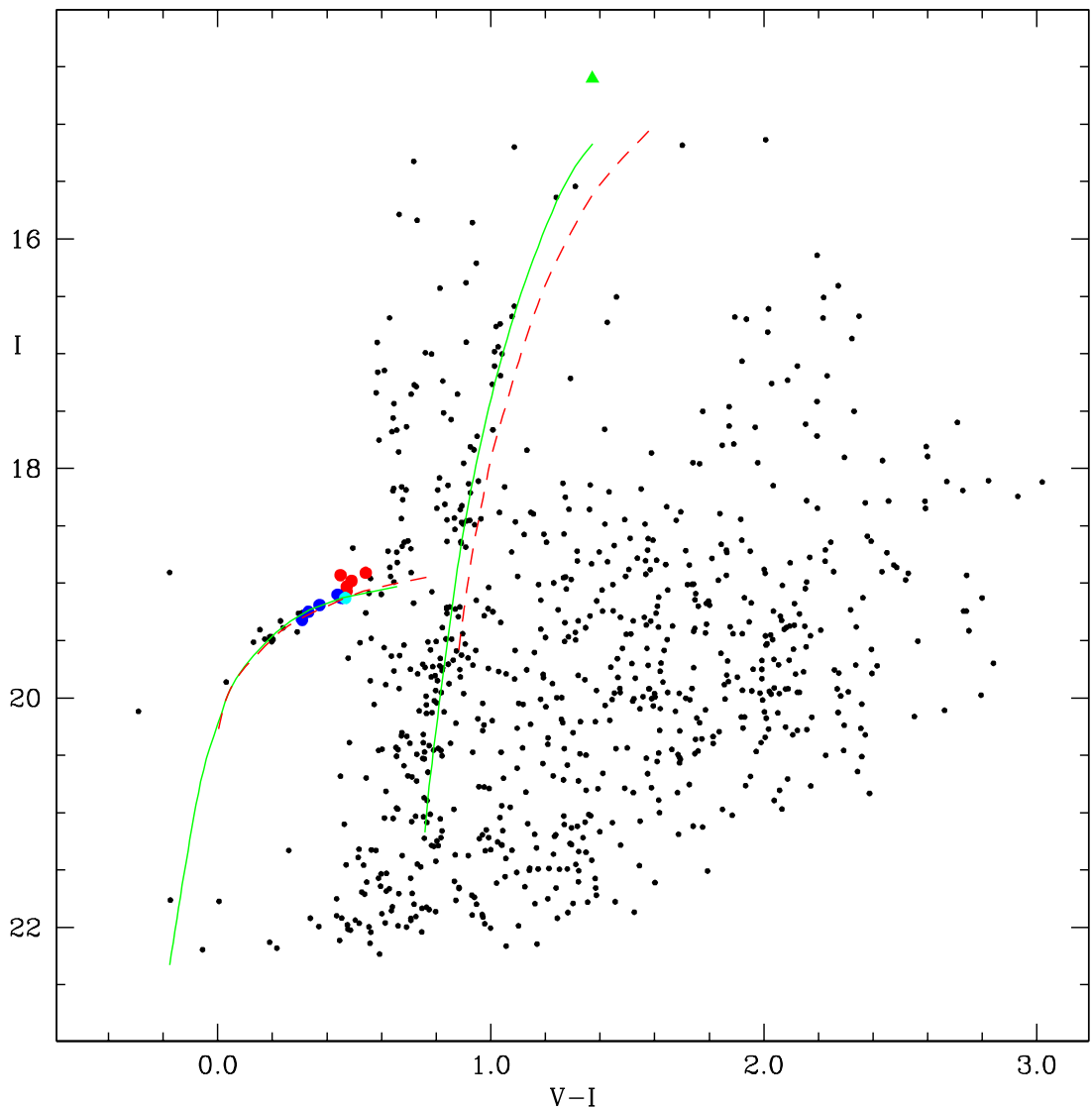}
\caption
{$I$, $V-I$ CMD of the Bo${\rm \ddot{o}}$tes 
dSph galaxy with the variable stars
plotted in different colors. Red dots are {\it ab-}type RR Lyrae stars  
(RRab) and the variable $\sim$ 0.2 mag above the galaxy HB; blue and cyan dots are first
overtone (RRc) and double mode (RRd) pulsators, respectively; the green triangle
is a long period variable (LPV). 
Solid and dashed lines show the mean ridge lines of M15 and M3, shifted in
magnitude and adjusted in reddening to fit the galaxy HB. \label{f:fig1}
}
\end{figure}

\clearpage

\begin{figure}
\plotone{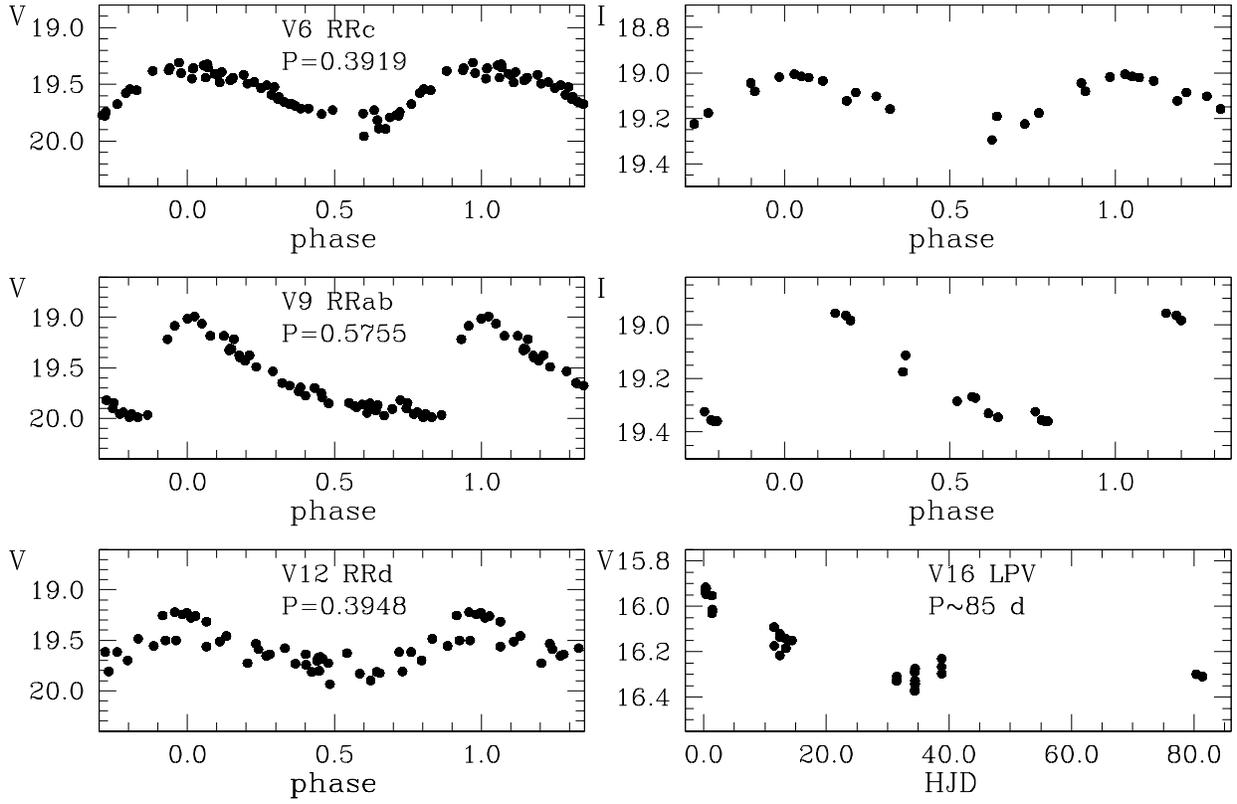}
\caption
{$V$ and $I$ light curves of variable stars in the Bo${\rm \ddot{o}}$tes 
dSph galaxy.
Top panels: {\it c-} and {\it ab-}type RR Lyrae stars. Bottom left:
$V$ light curve of the RRd star; bottom right: light curve in 
Heliocentric Julian Day (HJD) of the LPV.
\label{f:fig2}
}
\end{figure}

\clearpage
    
\begin{figure}
\plotone{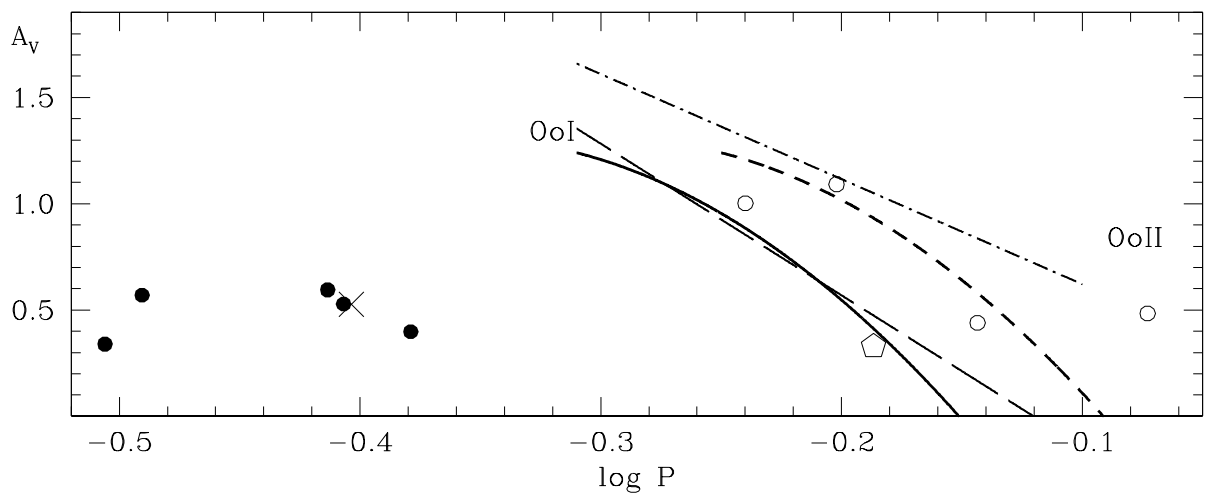}
\caption
{Period-Amplitude diagram in the $V$ band for the
Boo 
\rrl\ stars.  Filled and open circles are {\emph c-}
and {\emph ab-type} \rrl\ stars, respectively. The X sign is the double mode star.
The large pentagon is V5 the variable about 0.2 mag above the HB having rather small amplitude. 
The straight lines
are the positions of the Oosterhoff type I (OoI) and II (OoII)
Galactic GCs according to \citet{clement00}. Period-amplitude distributions of the
{\it bona fide} regular (solid curve) and well-evolved (dashed curve) {\it ab-}type \rrl\ stars in M3 
from \citet{cacciari05} are also shown for comparison.
\label{f:fig3}
}
\end{figure}

\clearpage

\begin{figure}
\plotone{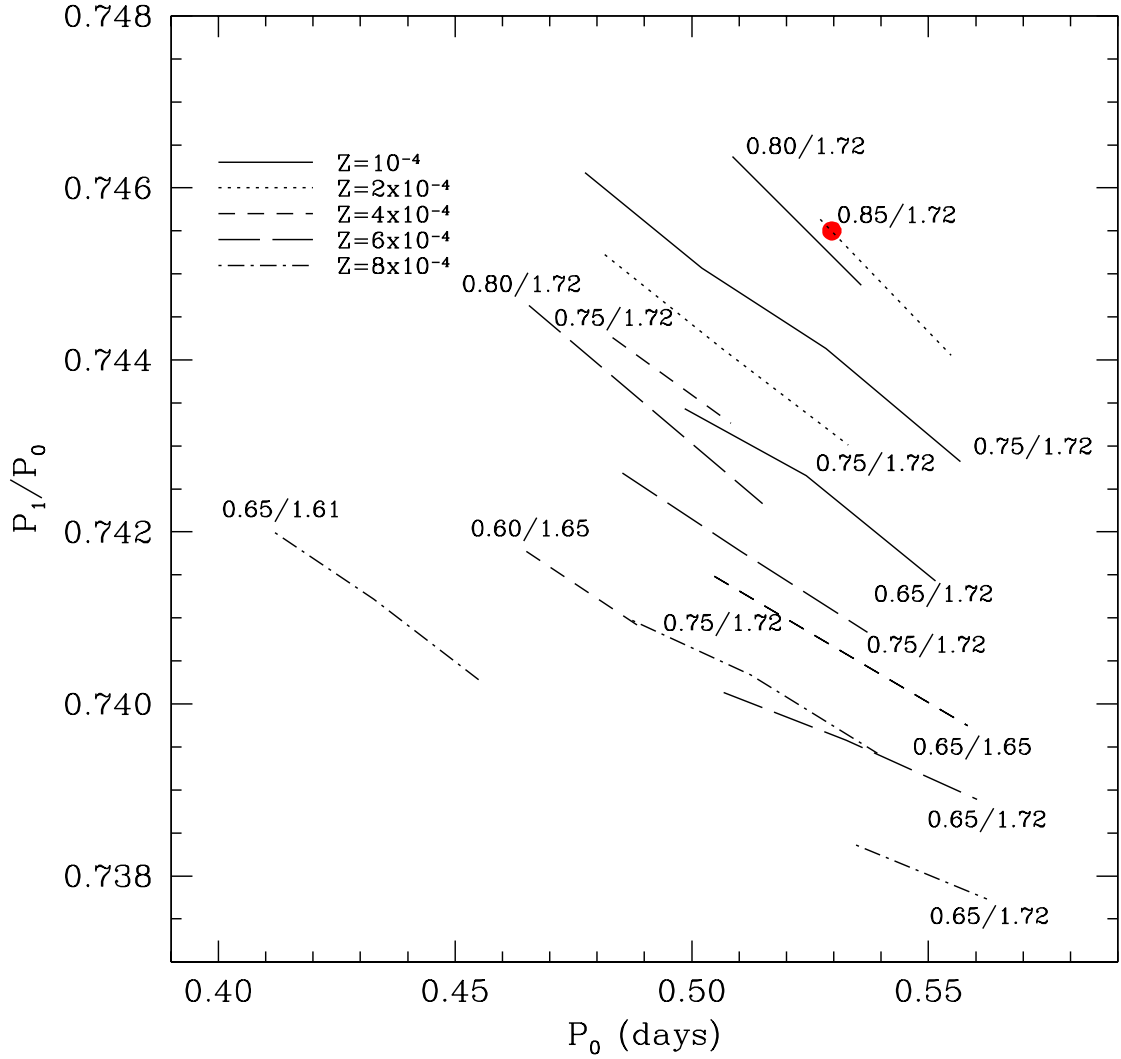}
\caption 
{Comparison between the observed location of the RRd star
V12  (filled circle) in the Petersen diagram, and the theoretical predictions
of \citet{bono96} and \citet{bragaglia01} pulsation models.
Numbers near each line indicate the model mass and luminosity. 
\label{f:fig4}
}
\end{figure}

\end{document}